\documentclass[aps,prx,twocolumn,superscriptaddress,amsmath,amssymb]{revtex4-1}

\usepackage{graphicx}
\usepackage{dcolumn}
\usepackage{bm}
\usepackage{chemformula} 
\usepackage{braket}
\usepackage{comment}
\usepackage{color}
\usepackage{hyperref}

\begin{document}

\title{A Jastrow wave function for the spin-1 Heisenberg chain: the string order revealed by the mapping to the classical Coulomb gas}

\author{Davide Piccioni}
\email[]{dpiccion@sissa.it}
\affiliation{Scuola Internazionale Superiore di Studi Avanzati (SISSA), Via Bonomea 265, I-34136, Trieste, Italy}

\author{Christian Apostoli}
\affiliation{Dipartimento di Fisica ``Aldo Pontremoli'', Universit\`a degli Studi di Milano, via Celoria 16, I-20133 Milano, Italy}

\author{Federico Becca}
\affiliation{Dipartimento di Fisica, Universit\`a di Trieste, Strada Costiera 11, I-34151 Trieste, Italy}

\author{Guglielmo Mazzola}
\affiliation{Institute for Computational Science, University of Zurich, Winterthurerstrasse 190, 8057 Zurich, Switzerland}

\author{Alberto Parola}
\affiliation{Dipartimento di Scienza e Alta Tecnologia, Universit\`a dell'Insubria, Via Valleggio 11, I-22100 Como, Italy}

\author{Sandro Sorella}
\affiliation{Scuola Internazionale Superiore di Studi Avanzati (SISSA), Via Bonomea 265, I-34136, Trieste, Italy}

\author{Giuseppe E. Santoro}
\affiliation{Scuola Internazionale Superiore di Studi Avanzati (SISSA), Via Bonomea 265, I-34136, Trieste, Italy}
\affiliation{International Centre for Theoretical Physics (ICTP), Strada Costiera 11, I-34151 Trieste, Italy}
\affiliation{CNR-IOM, Consiglio Nazionale delle Ricerche - Istituto Officina dei Materiali, c/o SISSA, Via Bonomea 265, I-34136 Trieste, Italy}

\date{\today}

\begin{abstract}
We show that a two-body Jastrow wave function is able to capture the ground-state properties of the $S=1$ antiferromagnetic Heisenberg chain 
with the single-ion anisotropy term, in both the topological and trivial phases. Here, the optimized Jastrow pseudo-potential assumes a very 
simple form in Fourier space, i.e., $v_{q} \approx 1/q^2$, which is able to give rise to a finite string-order parameter in the topological 
regime.  The results are analysed by using an exact mapping from the quantum expectation values over the variational state to the classical 
partition function of the one-dimensional Coulomb gas of particles with charge $q=\pm 1$. Here, two phases are present at low temperatures: 
the first one is a diluted gas of dipoles (bound states of particles with opposite charges), which are randomly oriented (describing the 
trivial phase); the other one is a dense liquid of dipoles, which are aligned thanks to the residual dipole-dipole interactions (describing 
the topological phase, with the finite string order being related to the dipole alignment). Our results provide an insightful interpretation 
of the ground-state nature of the spin-1 antiferromagnetic Heisenberg model.
\end{abstract}

\maketitle

\section{Introduction}\label{sec:intro}

Since the early days of quantum mechanics, the simulation of correlated quantum many-body systems has been an interesting problem in physics. 
Due to the paucity of exact results, many different approximation schemes have been proposed, most of them relying on some kind of variational 
{\em Ansatz}, including the variational quantum Monte Carlo approach, where the square modulus of a trial wave function is sampled by a Markov 
chain Monte Carlo algorithm~\cite{becca2017}. In the last few years, many efforts have been devoted to improving the variational description 
of correlated quantum many-body system, by various approaches, including tensor networks~\cite{verstraete2004,perez2007,orus2019} and 
machine-learning techniques {\em Ans\"{a}tze}~\cite{carleo2017,torlai2018}. However, these powerful approaches feature a large number of 
parameters, which may be difficult to optimize, and do not offer a transparent physical interpretation. By contrast, the Jastrow wave 
function~\cite{jastrow1955} has been introduced in 1955 to describe strongly-correlated systems on the continuum and found one of its first 
and most important applications in the description of the ground state of \ch{^{4}He}~\cite{mcmillan1965}. The Jastrow state is a simple 
variational {\em Ansatz}, including a correlation operator acting on an uncorrelated state, hence allowing for a clear physical interpretation 
of its parameters after the optimization. In its simplest definition (as used in the early applications~~\cite{mcmillan1965}), the correlation
factor involves two-body terms.

While it is obvious that modern {\em Ans\"{a}tze} can provide quantitatively better results compared to Jastrow states on several microscopic
systems (e.g., $S=1/2$ Heisenberg models~\cite{carleo2017}), it would be interesting to identify cases where these sophisticated trial states 
outperform qualitatively the simple Jastrow wave functions, i.e., where the application of a more accurate {\em Ansatz} predicts the emergence 
of new phases of matter. For example, accurate machine-learning inspired {\em Ans\"{a}tze} have successfully improved the energy of benchmark 
problems with known solutions, but only rarely provided additional understanding on contested phase diagrams. To provide physical insights on 
the ground-state properties of correlated models, they are often combined with more traditional approaches, like in the case of frustrated spin 
models in two and three dimensions, most notably the $J_1{-}J_2$ model on the square lattice~\cite{nomura2021} and the Heisenberg model on the 
pyrochlore lattice~\cite{astrakhantsev2021}.

The quantum $S=1$ antiferromagnetic Heisenberg chain, including the single-ion anisotropy term, provides a fair example of a simple model that
displays a rich phenomenology. The Hamiltonian is given by
\begin{equation}\label{eq:model}
\hat{\cal H} = J \sum_{j=1}^{L} \hat{\bf S}_{j} \cdot \hat{\bf S}_{j+1} + D \sum_{j=1}^{L} \big(\hat{S}^z_{j}\big)^2 \;,
\end{equation}
where $\hat{\bf S}_{j}=(\hat{S}^x_{j},\hat{S}^y_{j},\hat{S}^z_{j})$ is the spin-1 operator on the $j$-th site and periodic boundary conditions are 
assumed on chains with $L$ sites. The interest in the model with $D=0$ grew after Haldane proposed his conjecture about integer-spin chains being 
gapped~\cite{haldane1983}. Later on, an exact solution has been found by Affleck-Kennedy-Lieb-Tasaki (AKLT) for the case where a bi-quadratic 
interaction is added~\cite{affleck1987}, the ground state being gapped and unique with periodic boundary conditions, featuring exponentially 
decaying correlations, as predicted by Haldane for the Hamiltonian of Eq.~\eqref{eq:model} with $D=0$. Furthermore, it has been highlighted the 
existence of a ``hidden'' $Z_2\times Z_2$ symmetry~\cite{kennedy1992}, which is broken in this gapped ground state. More recently, the Haldane 
state has been recognized as an example of symmetry-protected topological (SPT) phase~\cite{pollmann2010}, with exponentially decaying spin-spin 
correlation functions and a non-local order parameter that is revealed by the so-called ``string operator''~\cite{dennijs1989,kennedy1992}:
\begin{equation}\label{eq:string_operator}
\hat{O}^{\mathrm{string}}_{i,j} \equiv \hat{S}^z_i \exp{\left\{i\pi \sum_{l=i+1}^{j-1} \hat{S}^z_l \right\}} \hat{S}^z_j \;.
\end{equation}
The non-local order corresponds to a finite average value $C^{\mathrm{string}}_{i,j} \equiv \braket{\hat{O}^{\mathrm{string}}_{i,j}}$, computed on 
the ground state, at large distance:
\begin{equation}\label{eq:string_correlation}
\lim_{|i-j|\to \infty} C^{\mathrm{string}}_{i,j} \neq 0 \;.
\end{equation}

The ground-state properties of the Hamiltonian~\eqref{eq:model} have been investigated using a large variety numerical techniques, including 
exact diagonalizations~\cite{chen2003}, density-matrix renormalization group (DMRG) approaches~\cite{hu2011,pollmann2013}, and stochastic-series 
expansion techniques~\cite{albuquerque2009,huang2021}. In addition, a variational Monte Carlo approach based on Gutzwiller-projected wave functions 
in the fermionic representation has been employed, showing a high accuracy in describing the ground state of $S=1$ Heisenberg model~\cite{liu2012}. 
Besides the Haldane phase, there are two trivial phases that are stabilized for large negative and positive values of $D/J$ (with $J>0$): the first 
one is a (doubly degenerate) N\'{e}el antiferromagnet that can be adiabatically connected to the Ising configuration where all spins align along 
the $z$-axis, i.e., on each site the $z$-component of the spin operator assumes values $m_{j}=\pm 1$; the second one is a spin-nematic phase, where 
the ground state is connected to a state in which the $z$-component of the spin vanishes on each site, i.e., $m_{j}=0$. The transition between the 
Haldane and the nematic phases is particularly interesting because it has a topological character, with a gapless critical point, which has been 
recognized to be a gapless Tomonaga-Luttinger liquid phase~\cite{huang2021}.

In this paper we show that, despite its apparent simplicity, a Jastrow wave function is able to describe correctly the ground state of the $S=1$ 
Heisenberg antiferromagnetic chain, including the Haldane phase. In the following, we will focus on the case with $D/J \geq 0$. Our research is 
motivated by the fact that, in the context of the one-band Hubbard model, it has been shown that an appropriate Jastrow factor is able to turn 
a gapless metal or superfluid into a gapped Mott insulator~\cite{capello2005,capello2007}. We also mention that a Jastrow wave function for 
generalized AKLT models has been proposed~\cite{arovas1988}, exploiting the analogy between the valence-bond states (within the Schwinger-boson
representation) and the Laughlin wave function for the fractional quantum Hall effect~\cite{laughlin1983}. Within spin models, the Jastrow 
{\em Ansatz} has been largely employed, mainly for $S=1/2$ cases~\cite{manousakis1991}. However, it appears rather non-trivial that a Jastrow 
wave function is able to represent a state with a finite string-order parameter. The issue is that the Jastrow wave function is parametrized by 
a two-body pseudo-potential $v_{r}$ that describes the spin-spin correlations at distance $r$ and, therefore, it is not obvious that it is able 
to capture many-body correlations, like the string correlation that includes the knowledge of {\em all} spin values between two (distant) sites. 
Instead, we show that it is actually possible to properly represent the Haldane phase, the string correlations arising thanks to the long-range 
character of the pseudo-potential $v_{r}$, i.e., $v_{q} \approx 1/q^2$ in Fourier space. 

Remarkably, the fully-optimized $v_{r}$ can be well approximated by using only two parameters, one regulating the density of non-zero $m_{j}$ 
and the other one defining the strength of correlations between (distant) pairs of spins, i.e., $m_{j}$ and $m_{j+r}$. Most importantly, assuming 
this simplified version of the pseudo-potential, we perform a mapping between the modulus squared of the Jastrow factor and the partition function 
of a {\em one-dimensional} classical Coulomb gas. Then, the properties of the variational state are related to (thermal) ones of the corresponding 
classical model. The use of similar mappings has proved to be very useful in strongly-correlated systems, the most notable example is given by the 
Laughlin state for the fractional quantum Hall effect~\cite{laughlin1983}. On the lattice, a similar approach has been used for the density-density 
Jastrow factor, to investigate the resulting metal-insulator (Mott) transition~\cite{capello2006}. In the present context, the classical model is 
a Coulomb gas of particles with charge $q=\pm 1$. Here, two phases are present at low temperatures (relevant for the actual pseudo-potential). The 
first one is described by a diluted gas of dipoles with $m_{j}=\pm 1$, on top of a background with $m_{j}=0$; these dipoles are randomly oriented, 
since thermal fluctuations overcome the residual dipole-dipole interaction. This phase corresponds to the nematic phase with large values of $D/J$. 
The second one contains a dense liquid of dipoles, which are aligned (hence, ordered) in space. Here, a finite string order is present, leading to 
a correct description of the Haldane phase for small values of $D/J$. The transition between these two phases is first order at low temperatures, 
with a jump in the charge density and a string-order parameter that also jumps from zero to a finite value; instead, at larger temperatures, the 
transition changes nature, with no discontinuity in the charge density and a string-order parameter that varies continuously from zero to a finite 
value.

The results obtained by using the simplified version of the pseudo-potential give an insightful vision of the ground-state nature of the $S=1$
Heisenberg model. However, on finite clusters, close to the transition between the topological and trivial phases, there is a small region where
the functional form of $v_{r}$ deviates from its simplified version, showing a less singular behavior at small-$q$ values $v_{q} \approx 1/q$. 
Here, robust spin-spin correlations are present in the $x{-}y$ plane, suggesting the existence of long-range order and gapless excitations.
However, by increasing the number of sites, the width of this region shrinks, suggesting that this ``intermediate'' phase disappears in the
thermodynamic limit and the transition point can be described by the classical Coulomb gas picture.

The paper is organized as follows: In section~\ref{sec:wavefun}, we introduce the variational Jastrow wave function; in section~\ref{sec:results},
we show the numerical results and discuss the mapping to the classical Coulomb gas model, which is pivotal to highlight the physical properties
of the Jastrow {\em Ansatz}; finally, in section~\ref{sec:concl}, we draw our conclusions.

\begin{figure}
\includegraphics[width=\columnwidth]{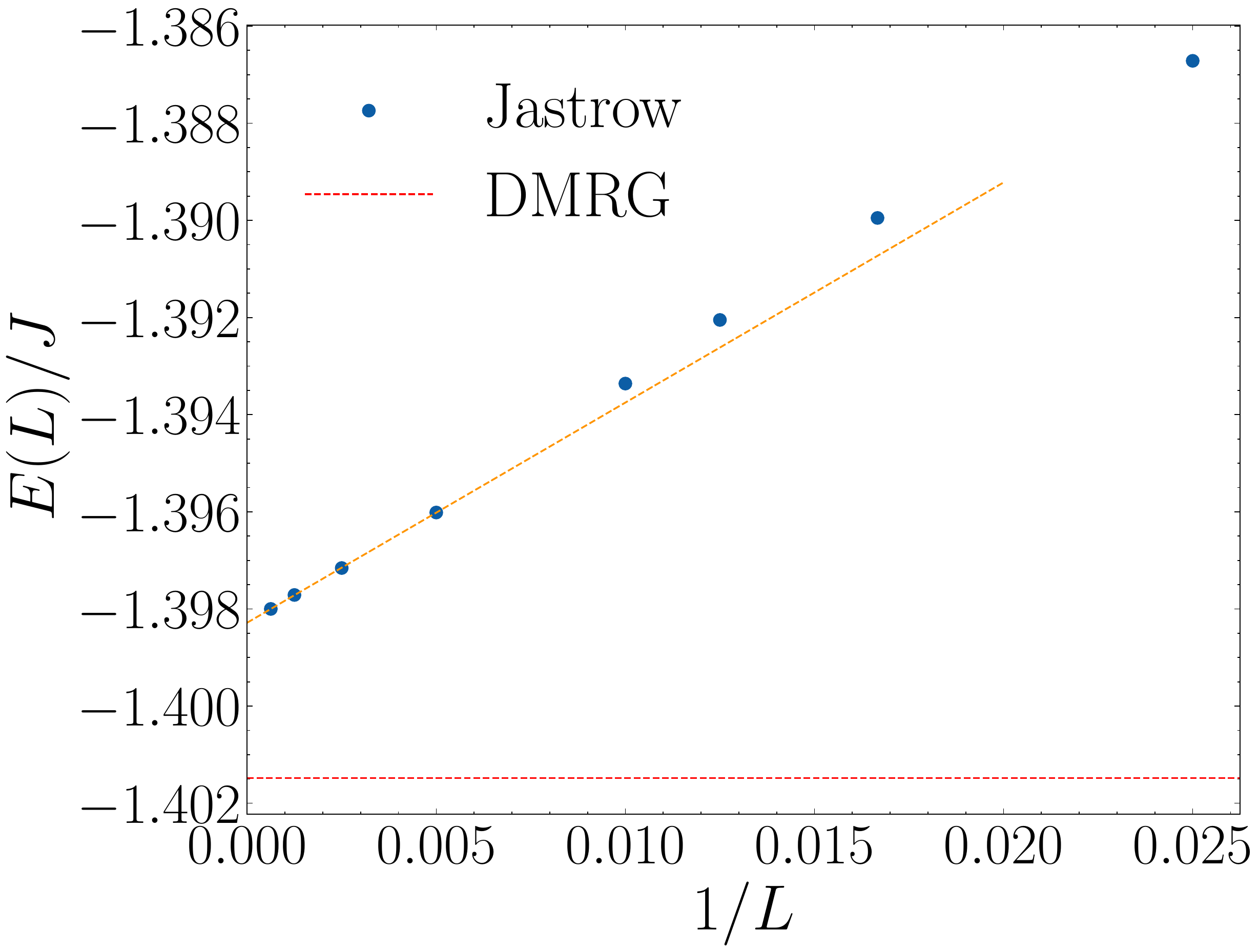}
\caption{\label{fig:energy} 
Energy per site $E(L)/J$ as a function of $1/L$ for the Jastrow {\em Ansatz} of Eq.~\eqref{eq:wavefunction} with optimized parameters for $D=0$. 
The red dashed line corresponds to the thermodynamic value obtained within DMRG, i.e., $E_{\mathrm{DMRG}}/J=-1.401484$,~\cite{white1993} while a 
linear extrapolation of the energies for the largest sizes gives $E(L\to \infty)/J=-1.3981(1)$.}
\end{figure}

\begin{figure*}
\centering
\includegraphics[width=\columnwidth]{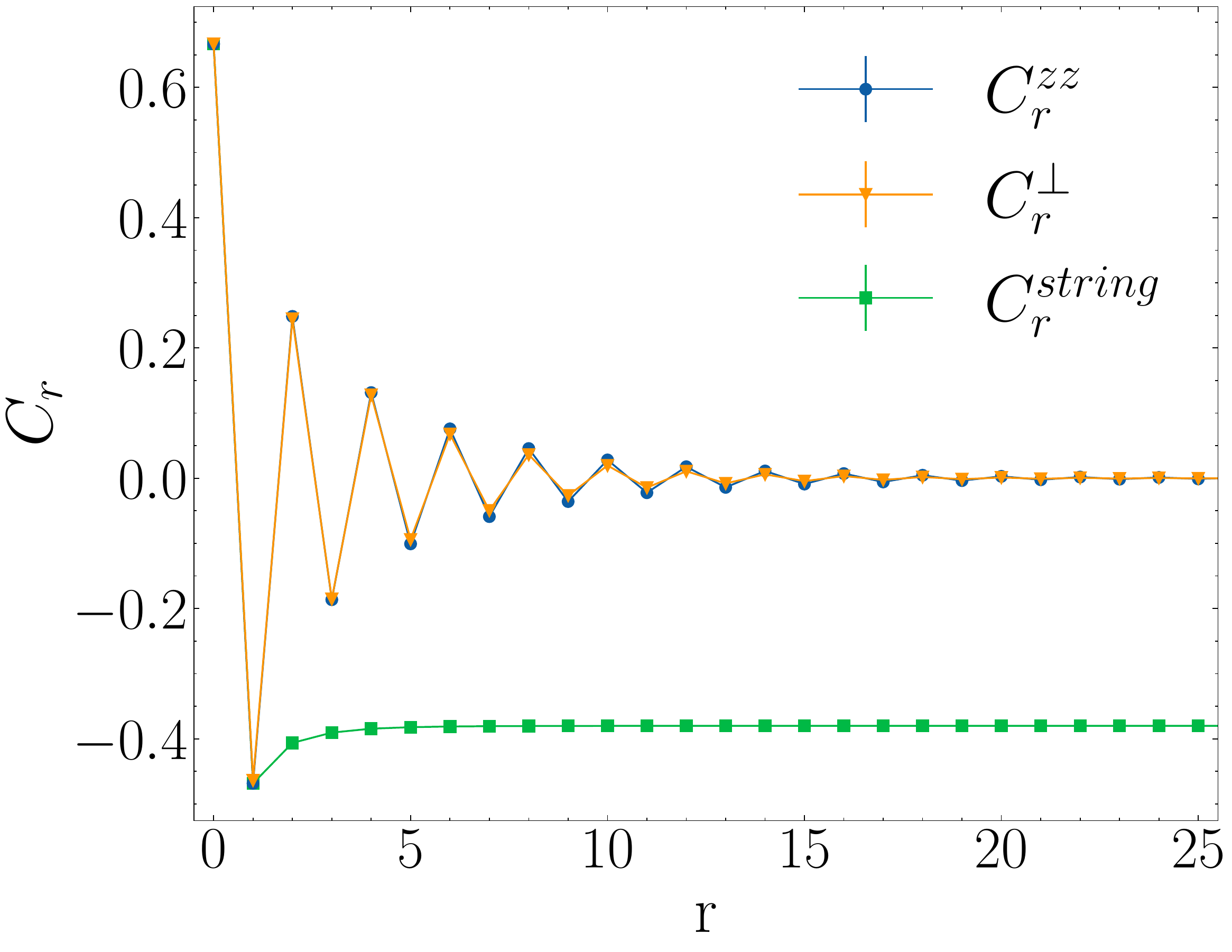}
\includegraphics[width=\columnwidth]{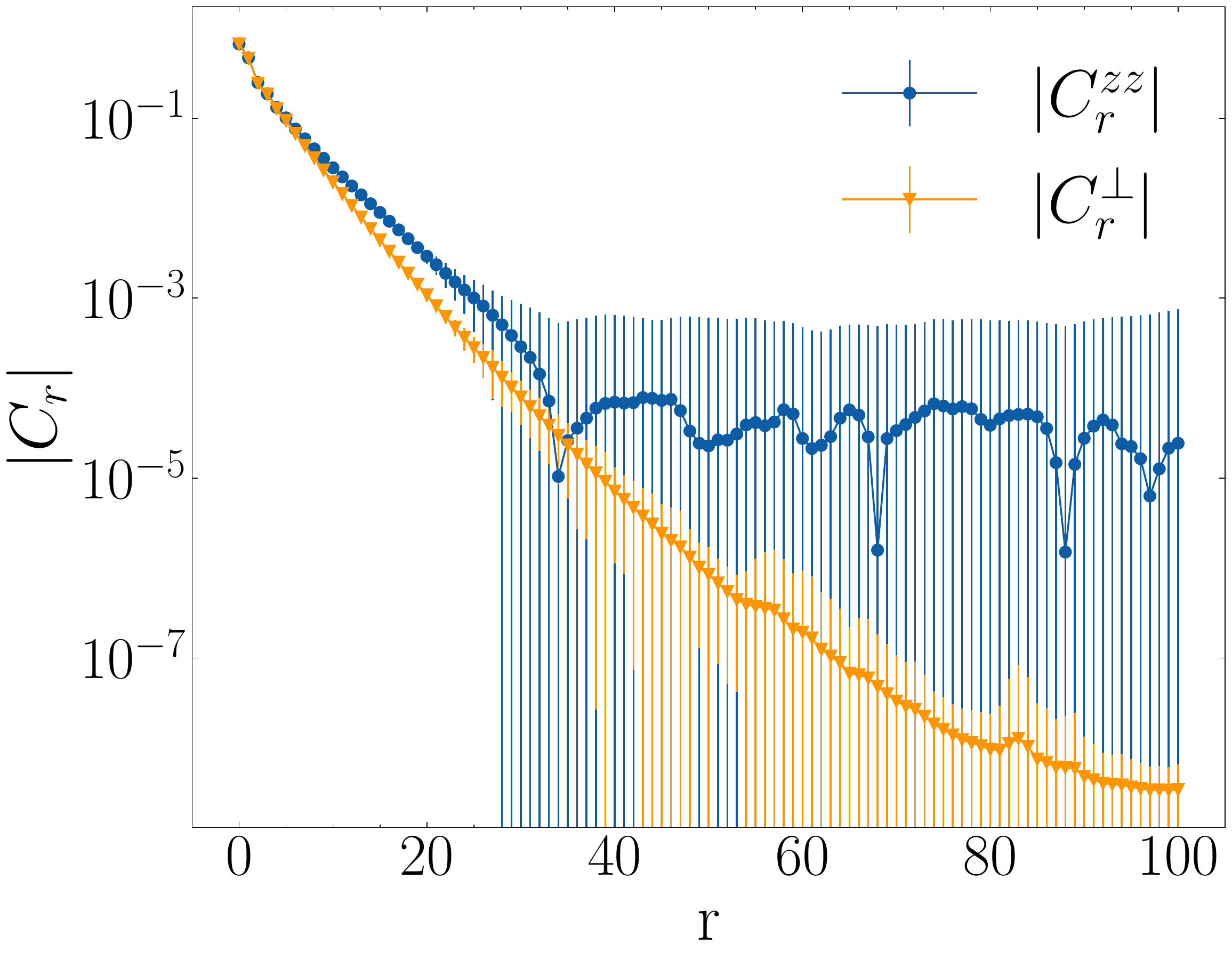}
\caption{\label{fig:correlations} 
Correlation functions for the optimized Jastrow state $D=0$ and $L=200$. (a) Spin-spin correlations $C^{zz}_{r}$ and $C^{\perp}_{r}$ and string 
correlations $C^{\mathrm{string}}_{r}$ [Eqs.~\eqref{eq:czz},~\eqref{eq:cxy}, and~\eqref{eq:string}] as a function of $r$. (b) Absolute value of 
the correlations $C^{zz}_{r}$ and $C^{\perp}_{r}$ in semi-log plot: their behavior is consistent with an exponential decay to zero.}
\end{figure*}

\section{The Jastrow wave function}\label{sec:wavefun}

Even though the exact ground-state wave function of the Hamiltonian~\eqref{eq:model} is not known on large clusters, its sign structure obeys the
so-called Marshall sign rule~\cite{marshall1955}. Indeed, the model is defined on a bipartite lattice, which can be divided into two sub-lattices
$A$ and $B$ with off-diagonal terms only connecting different sublattices. In addition, the ground state belongs to the sector with zero total 
magnetization, i.e., $\hat{M}=\sum_{j} \hat{S}^z_{j}=0$. Then, by defining the basis $\{ \ket{x} \}$ in which the $z$-component of the spin 
operator $m_{j}$ is given on each site (and $M=\sum_{j} m_{j}=0$), the ground state is written as:
\begin{equation}\label{eq:marshall_peierls}
\ket{\Psi_0} = \sum_{x} (-1)^{N(x)} f(x)\ket{x} \;,
\end{equation}
where $f(x)>0$ is the (unknown) amplitude and $N(x)=\sum_{j\in B}[1+m_{j}(x)]$, where $m_{j}(x)$ is value of $\hat{S}^z_{j}$ in $\ket{x}$, i.e., 
$\hat{S}^z_{j} \ket{x} = m_{j}(x) \ket{x}$.

The Jastrow wave function is obtained by applying a (positive) Jastrow factor $\hat{\mathcal J}$ to a state $\ket{\Phi_0}$ that encodes the 
Marshall sign rule (within the subspace with $M=0$):
\begin{equation}\label{eq:wavefunction}
\ket{\Psi_J}=\hat{\mathcal J} \ket{\Phi_0} \;.
\end{equation}
In our case, we take
\begin{equation}\label{eq:democratic}
\ket{\Phi_0}= \hat{\mathcal P}_{M=0} \prod_{j} \frac{1}{\sqrt{3}} \left [ (-1)^j \ket{0}_{j} + \ket{+1}_{j} + \ket{-1}_{j} \right ] \;,
\end{equation}
where $\hat{\mathcal P}_{M=0}$ is the projector into the subspace with $M=0$ and $\ket{0}_{j}$ and $\ket{\pm 1}_{j}$ are the eigenkets of 
$\hat{S}^z_{j}$ with eigenvalue $0$ and $\pm 1$, respectively. As a result, $\ket{\Phi_0}$ is a linear combination of all possible states with 
vanishing magnetization and equal weights (the signs of the spin configurations being consistent with the ones of the Marshall-sign rule). We 
remark that $\ket{\Phi_0}$ resembles the N\'{e}el antiferromagnetic state pointing in the $x$-direction, where the local states are given by 
$1/2 \left [ (-1)^j \sqrt{2} \ket{0}_{j} + \ket{+1}_{j} + \ket{-1}_{j} \right ])$.

The Jastrow factor has the following form:
\begin{equation}\label{eq:jastrow}
\hat{\mathcal J} = \exp{\left\{\frac{1}{2}\sum_{i,j} v_{i,j} \hat{S}^z_i \hat{S}^z_j\right\}} \;,
\end{equation}
where $v_{i,j}$ defines the so-called pseudo-potential, which, in translationally invariant cases, depends only upon the relative distance between 
the two sites $i$ and $j$: $v_{i,j}=v_{|i-j|}$. The correct physical properties are obtained when all the independent distances (e.g., $L/2$ on 
chains with periodic boundary conditions) are optimized independently. The Jastrow pseudo-potential is optimized minimizing the variational energy 
of the $S=1$ Hamiltonian~\eqref{eq:model} by using the stochastic reconfiguration technique~\cite{sorella2005}. An excellent approximation can be 
obtained by a simple parametrization, which includes only two parameters, one for the on-site term and the other one for the long-range behavior 
(see below). This simplified form of the pseudo-potential has a smooth convergence in the thermodynamic limit in Fourier space and is pivotal to 
give an insightful interpretation of the physical properties. In this case, the two parameters can be easily optimized within standard minimization
approaches.

\section{Results}\label{sec:results}

Here, we discuss the numerical results obtained by the Jastrow wave function~\eqref{eq:wavefunction} for the generalized Heisenberg Hamiltonian 
of Eq.~\eqref{eq:model}. In addition, we also present the mapping to the classical partition function and the corresponding phase diagram. 

\subsection{The Haldane phase at $D=0$}\label{sec:haldane_phase}

Let us begin by considering the isotropic Heisenberg point, with $D=0$, corresponding to the Haldane phase, which is gapped and presents a 
non-local order parameter~\eqref{eq:string_correlation}. First of all, in order to check how well the Jastrow {\em Ansatz} is able to reproduce 
the ground state of the spin-1 Heisenberg model, we compute the variational energy per site $E(L)$ for various sizes and compare it with the 
thermodynamic extrapolation obtained by the DMRG technique~\cite{white1993}, see Fig.~\ref{fig:energy}. For $L=200$, the relative error is 
smaller than $0.5\%$ and gets lower when the size of the system increases. By a linear extrapolation of the large-cluster results, we find that 
the thermodynamic energy of the Jastrow state is $E(L\to \infty)/J=-1.3981(1)$ (i.e., the relative error with respect to DMRG is about $0.2\%$).

Remarkably, the Jastrow wave function is able to reproduce the Haldane-phase physics, exhibiting non-vanishing string order and exponentially 
decaying correlation functions. To show this fact, we report the spin-spin correlation functions:
\begin{eqnarray}\label{eq:czz}
C^{zz}_{r} &\equiv& \frac{1}{L} \sum_{j} \braket{\hat{S}^z_{j} \hat{S}^z_{j+r}} \;, \\
C^{\perp}_{r} &\equiv& \frac{1}{2L} \sum_{j} \braket{ \left ( \hat{S}^x_{j} \hat{S}^x_{j+r} + \hat{S}^y_{j} \hat{S}^y_{j+r} \right )}  \;,
\label{eq:cxy}
\end{eqnarray}
and the string correlations of the operator~\eqref{eq:string_operator}
\begin{equation}\label{eq:string}
C^{\mathrm{string}}_{r}= \frac{1}{L} \sum_{j} \braket{\hat{O}^{\mathrm{string}}_{j,j+r}} \;.
\end{equation}
Notice that, since the Jastrow wave function generically breaks the spin SU(2) symmetry, the spin-spin correlations along the $z$ axis and in the 
$x{-}y$ plane are different, even though the discrepancy is small. The results for $L=200$ are reported in Fig.~\ref{fig:correlations}. Here, both
$C^{zz}_{r}$ and $C^{\perp}_{r}$ are consistent with an exponential decay with distance, while $C^{\mathrm{string}}_{r}$ reaches a constant value, 
hinting at the presence of string order.

\begin{figure}
\includegraphics[width=\columnwidth]{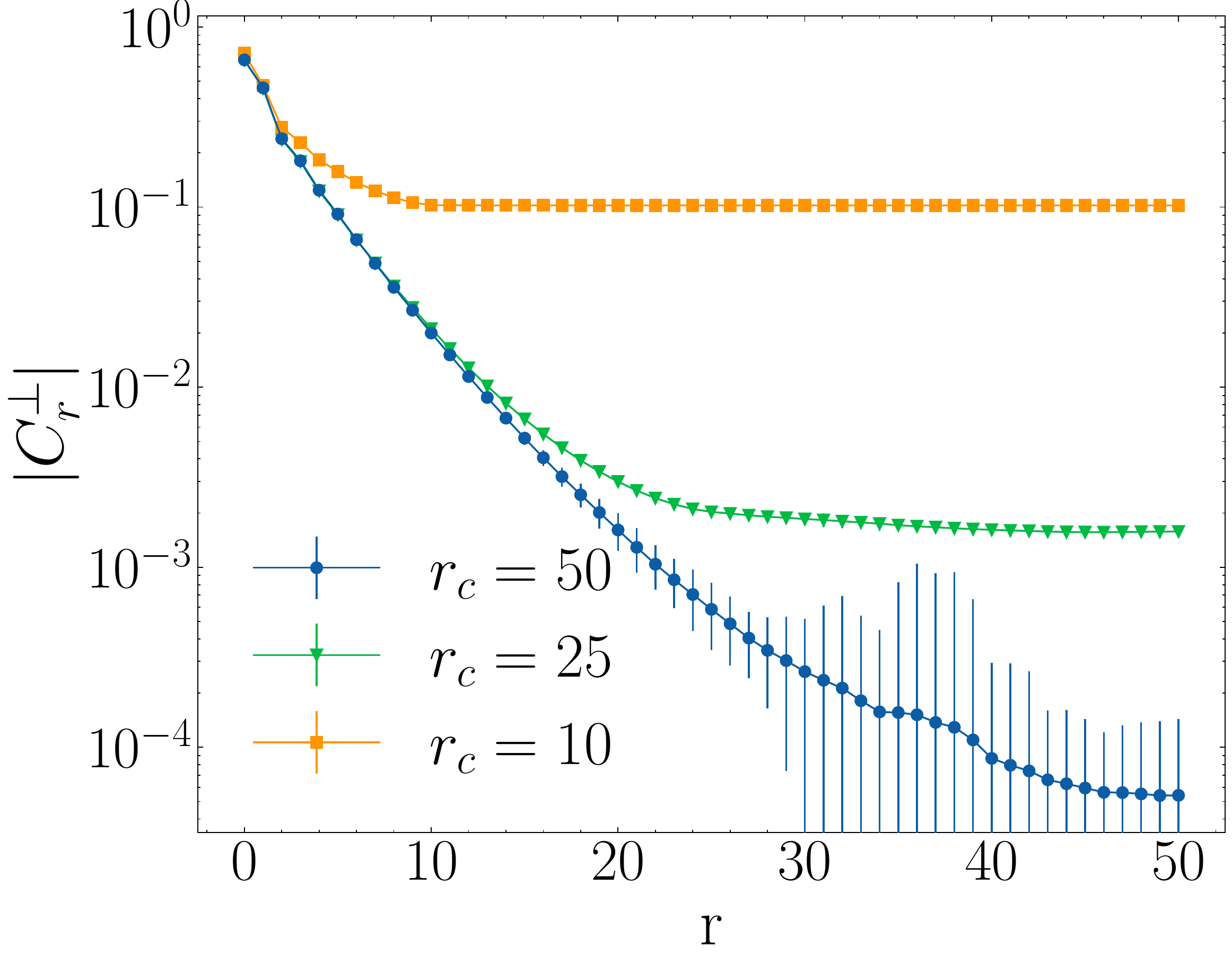}
\caption{\label{fig:vqshort} 
Spin-spin correlation $|C^{\perp}_{r}|$ in the log scale for the Heisenberg chain of $L=100$ sites at $D=0$ for different optimized Jastrow 
states, where a cutoff in the in the psuedo-potental is considered, i.e., $v_r=0$ for $r>r_{c}$. The cases with $r_c=25$ and $10$ are reported, 
as well as the case with no cutoff (i.e., $r_{c}=50$). The short-range character of the Jastrow wave function does not allow $|C^{\perp}_{r}|$ 
to decay exponentially to zero for $r \gtrsim r_{c}$.}
\end{figure}

Importantly, the optimized pseudo-potential has a long-range tail, which is crucial to reproduce the correct physical behavior of the Haldane phase. 
In particular, the optimized pseudo-potential $v_{q}$ in Fourier space behaves as $v_q \approx 1/q^2$ for small $q$, see Fig.~\ref{fig:vq}, where 
we report $v_q \times q^2$ for various sizes of the chain. The fact that a short-range Jastrow factor is not suitable to describe the Haldane phase 
is clear by optimizing $v_{r}$ only for distances $r<r_{c}$ (i.e., fixing $v_{r}=0$ for $r \ge r_{c}$). In Fig.~\ref{fig:vqshort}, the results of
the $x{-}y$ spin-spin correlations are reported for $r_{c}=10$ and $25$ on $L=100$, in comparison to the long-range case with $r_{c}=50$. The
former two cases clearly give a finite value of $C^{\perp}_{r}$ for $r \gtrsim r_{c}$. Furthermore, the string correlations are suppressed,
indicating that no string order is possible whenever the Jastrow factor is not long range.

\subsection{The large $D/J$ limit}\label{sec:largeD}

\begin{figure}
\includegraphics[width=\columnwidth]{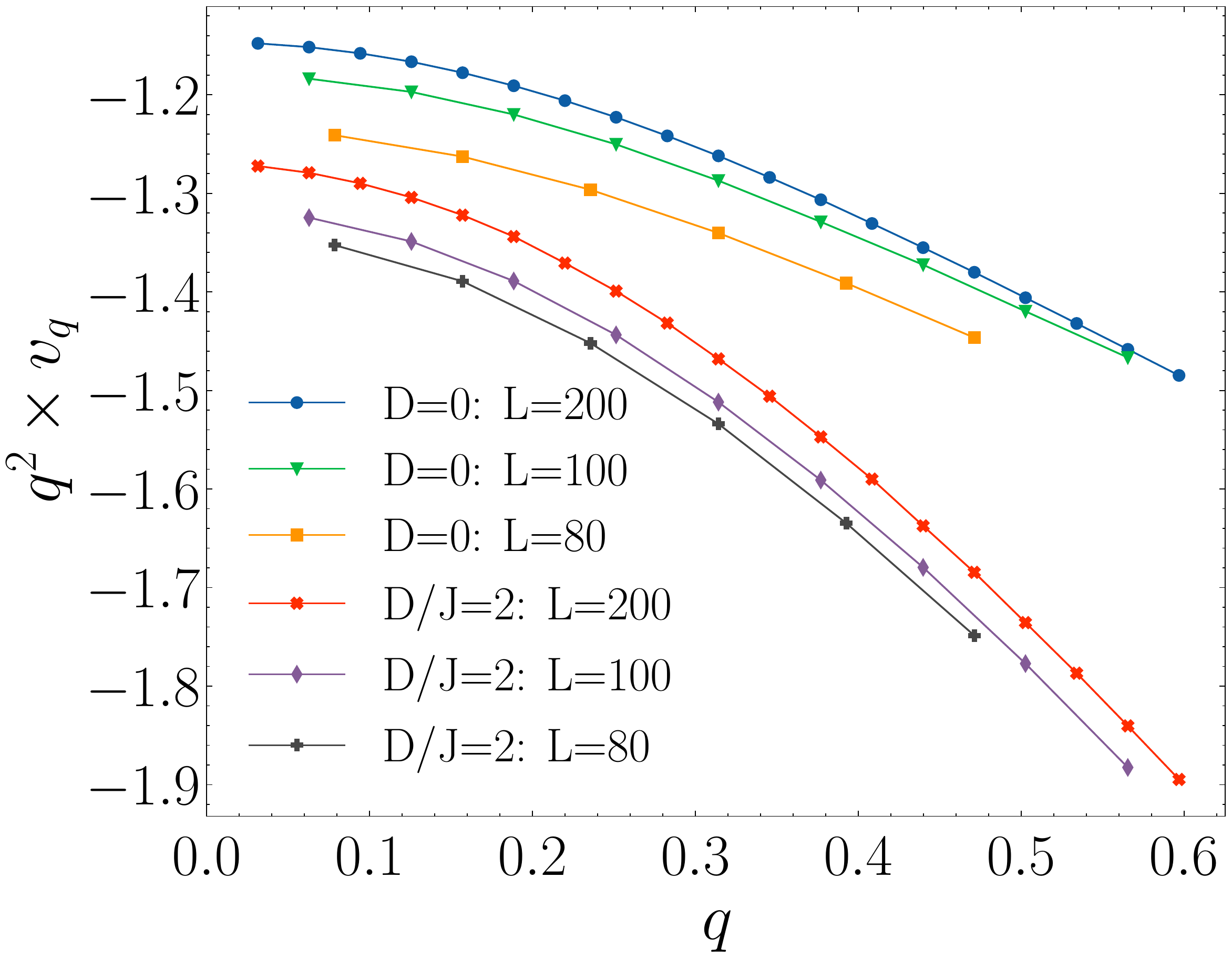}
\caption{\label{fig:vq} 
The optimized Jastrow pseudo-potential in Fourier space $v_q$ multiplied by $q^2$ as a function of $q$ for the Heisenberg model with $D=0$ and
$D/J=2$ for different values of $L$. The results show that $v_q \approx 1/q^2$ for $q \to 0$.}
\end{figure}

We now briefly discuss the regime where $D/J \gg 1$. It turns out that the the Jastrow wave function~\eqref{eq:wavefunction} correctly describes 
also this case. Here, the spin-spin correlations are still exponentially decaying with the distance. However, in contrast to the Haldane phase, 
the string correlations are correctly vanishing for large distance (not shown). Also in this case, the Jastrow pseudo-potential $v_{r}$ has a 
long-range tail, with $v_{q} \approx 1/q^2$ in the small-$q$ limit (see Fig.~\ref{fig:vq}), as in the Haldane regime. 

The fact that the same kind of behavior in the Jastrow pseudo-potential may give rise to both topological (with finite string order parameter)
and trivial (with no string order parameter) phases requires a deeper investigation, which is the subject of the next subsection.

\subsection{The classical Coulomb gas mapping}\label{sec:mapping}

In order to gain some understanding from the Jastrow wave function, we start by noticing that quantum expectation values over a given variational
state can be directly interpreted as thermal expectation values over classical probabilities. The idea underlying the classical mapping is that 
any quantum average of an operator $\hat{\cal O}$ over a state $\ket{\Psi}$ can be expressed as a sum over the complete basis set $\ket{x}$:
\begin{equation}\label{eq:mapping}
\braket{\hat{\cal O}} = \sum_{x} \frac{|\braket{x|\Psi}|^2}{\braket{\Psi|\Psi}} \frac{\braket{x|\hat{\cal O}|\Psi}}{\braket{x|\Psi}}
= \sum_{x} P(x) {\cal O}(x) \;,
\end{equation}
such that the wave function (squared) can be interpreted as a classical ``Boltzmann'' weight $P(x)$:
\begin{equation}\label{eq:boltzmann}
P(x) = \frac{1}{\cal Z} e^{-\beta {\cal E}(x)} \;,
\end{equation}
where $\beta {\cal E}(x) = -2 \log |\braket{x|\Psi}|$ defines an ``effective'' classical temperature $T=1/\beta$ and a classical energy 
${\cal E}(x)$. The quantum expectation value (at zero temperature) is then obtained by averaging the appropriate observable ${\cal O}(x)$ over 
$P(x)$. 

In the present scenario, $\braket{x|\Psi}$ is fully determined by the Jastrow factor, since $\braket{x|\Phi_0}$ of Eq.~\eqref{eq:democratic} is
constant (apart from the Marshall sign, which is irrelevant because of the modulus squared). The crucial point is that the fully-optimized Jastrow 
pseudo-potential $v_{i,j}$ may be accurately approximated by using only two parameters ($\beta$ and $\mu$):
\begin{equation}\label{eq:vqsimple}
v_{i,j} = \beta \left [ \mu \delta_{i,j} - \frac{1}{2L} \left ( d_{i,j}-\frac{L}{2} \right )^2 \right ] \;,
\end{equation}
where $d_{i,j} = \min \{ |i-j|, L-|i-j| \}$ is the distance between two sites $i$ and $j$. Therefore, the classical energy ${\cal E}(x)$ acquires
a particularly simple form in terms of $\{ m_{1}, \dots, m_{L} \}$ in the configuration $\ket{x}$ (here, for simplicity, we drop the label $x$ in 
$m_{j}$), namely ${\cal E}(x) \to E( \{ m_1, \dots, m_L \})$ with
\begin{equation}\label{eq:clen}
E(\{ m_1, \dots, m_L \}) = - \mu \sum_{j} m_j^2 + \frac{1}{2}\sum_{i,j} U_{i,j} m_i m_j \;,
\end{equation}
This expression of the classical energy identifies a one dimensional lattice gas, where each site may be either empty ($m_j=0$) or occupied by a 
unit charge $m_j=\pm 1$. The interaction potential between charges is given by 
\begin{equation}
U_{i,j}=\frac{1}{L} \left ( d_{i,j}-\frac{L}{2} \right )^2 \;,
\end{equation}
while the constraint of vanishing total magnetization $\sum_j m_j=0$ implies charge neutrality, which can be easily enforced by adding to the 
classical energy~\eqref{eq:clen} a term $\Lambda (\sum_{j} m_{j})^2$ in $U_{i,j}$, with $\Lambda \to \infty$. This constraint can be enforced by 
a shift in the interaction potential: $U_{i,j} \to U_{i,j} +\Lambda$, whose Fourier transform acquires a particularly simple form:
\begin{equation}\label{eq:defvq}
U_{q}=
\begin{cases}
\frac{1}{2[\sin (q/2)]^2} & {\rm for} \,\, q\neq 0 \;, \cr
\cr
L \Lambda + \frac{L^2+2}{12} & {\rm for} \,\, q=0 \;, \cr
\end{cases}
\end{equation}
which indeed corresponds to a Coulomb-like potential. Thus the resulting model describes a classical Coulomb gas of particles with charges $q=\pm 1$ 
in one spatial dimension.

The net consequence of this reasoning is that the quantum expectation values can be rewritten in terms of a classical model, whose partition 
function is 
\begin{equation}\label{eq:partition}
{\cal Z} = \sum_{ \{ m_{j} \}} e^{-\beta E(\{ m_1, \dots, m_L \})} \;.
\end{equation}
The physical properties of this classical model are far from being trivial and depend upon the values of the effective temperature $T=1/\beta$ 
and of chemical potential $\mu$, which fixes the average number of charges $m_{j}=\pm 1$.

\begin{figure}
\includegraphics[width=\columnwidth]{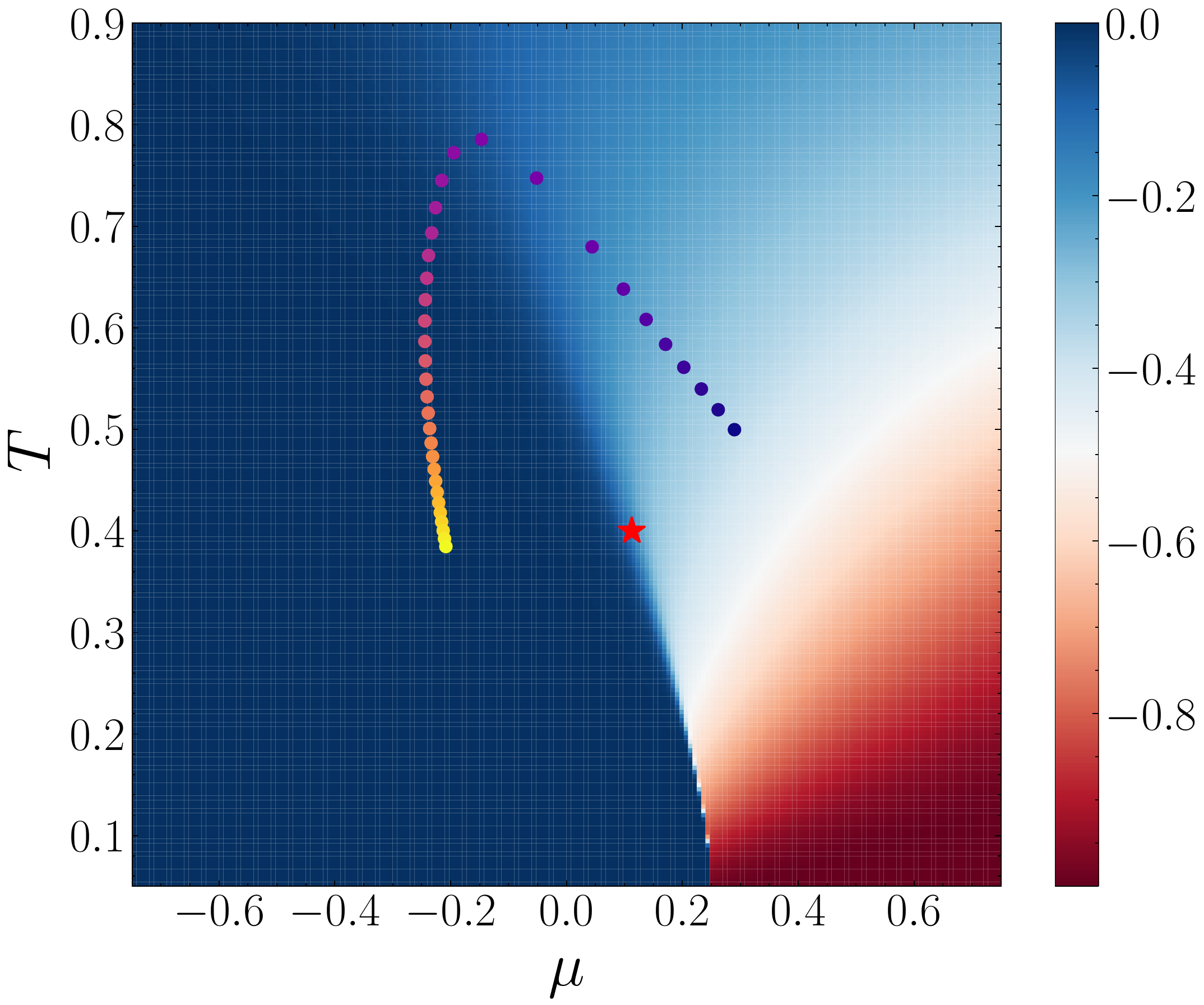}
\caption{\label{fig:phase_diag_classic_String} 
Phase diagram for the one-dimensional classical Coulomb gas as a function of $\mu$ and $T=1/\beta$ for a system with size $L=200$. The color-map 
represents the value of the string correlation at the maximum distance; it is easy to recognize the two phases, only the right one displaying 
string order. The points represented on top of the plot show the optimal parameters $\beta$ and $\mu$ for the variational wave function with the 
simplified Jastrow pseudo-potential, Eq.~\eqref{eq:vqsimple}, for a chain of $L=200$ sites. The values of $D/J$ span from $0$ (violet) to $3.4$ 
(yellow). A red star marks the tricritical point where, in the thermodynamic limit, the transition between the phases passes from first order to second order.}
\end{figure}

\begin{figure*}
\centering
\includegraphics[width=\columnwidth]{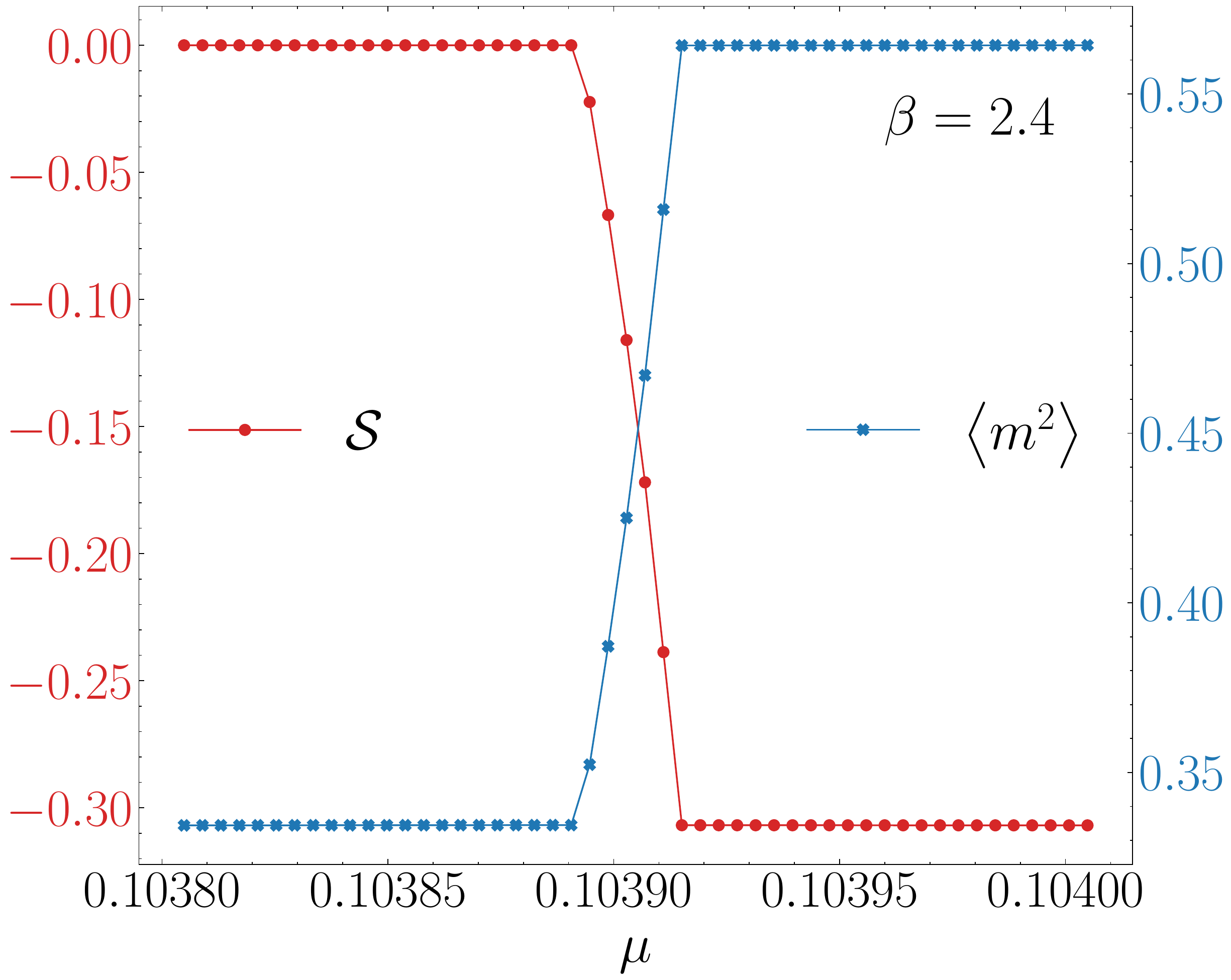}
\includegraphics[width=\columnwidth]{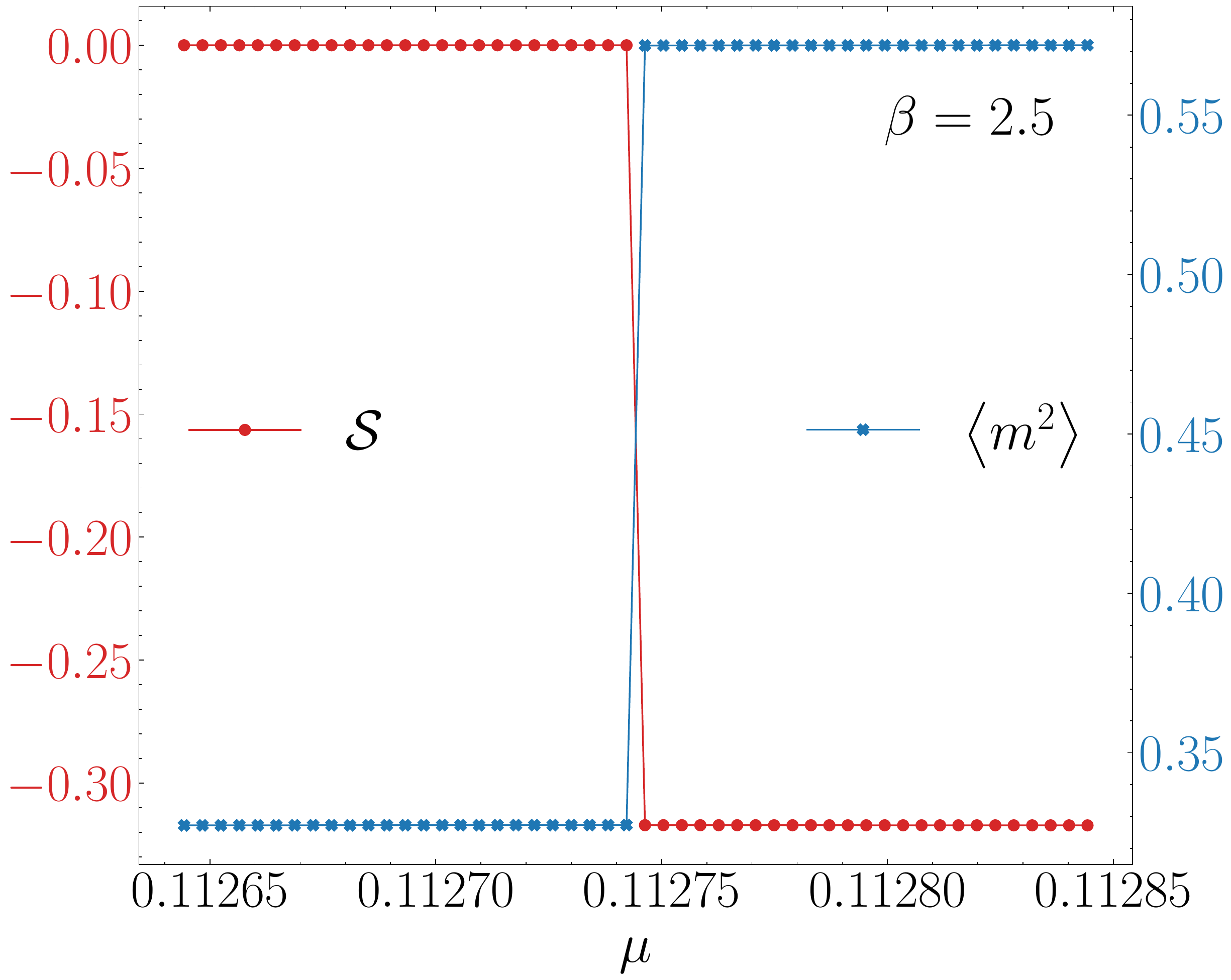}
\caption{\label{fig:string_density} 
Density $\braket{m_j^2}$ and string order $\mathcal{S}$ evaluated in the thermodynamic limit as functions of $\mu$ for two different values of 
$\beta$; in both cases, the scale over which $\mu$ varies is very small. (a) For $\beta=2.4$, the transition between the two phases is second 
order. (b) For $\beta=2.5$, the transition is first order.}
\end{figure*}

After some straightforward but non-trivial algebra (see Appendix~\ref{app1}), the partition function can be written in terms of an integral
over a continuous variable $Q$ and the trace of a tridiagonal symmetric matrix $\mathbb{T}(Q)$, whose elements are given by 
\[
T_{s,s^\prime}(Q)=
t(Q,s,s^\prime) \left [ \delta_{s,s^\prime}+ e^{\beta\mu} \delta_{s,s^\prime+1}+ e^{\beta\mu} \delta_{s,s^\prime-1}\right ] \;,
\]
where 
$t(Q,s,s^\prime)=e^{-\frac{\beta}{2} \left [ (Q+s)^2+(Q+s^\prime)^2\right ]}$:
\begin{equation}\label{eq:partitionfinal}
{\cal Z} = \sqrt{\frac{\beta L}{\pi}} \, \int^{1/2}_{-1/2} dQ \,{\rm Tr}\,\left [ \mathbb{T}(Q)\right ]^L \;.
\end{equation}
As previously noted [see Eq.~\eqref{eq:mapping}], the quantum expectation value (at zero temperature) of an operetor $\hat{\cal O}$ is obtained 
by averaging the appropriate observable ${\cal O}(x)$ over $P(x)$. The matrix elements ${\cal O}(x)$ can be easily handled if the corresponding 
operator $\hat{\cal O}$ is expressed in terms of the $z$ component of the local spins $\hat S_j^z$. In this case ${\cal O}(x)$ just becomes a 
function of the classical occupation numbers $m_j$ and the average acquires a transparent meaning, like, for example, the case of the local 
density of charges $\braket{m_j^2}$. Following the same procedure detailed in Appendix~\ref{app1}, the average density is shown to be given in 
terms of the above matrix $\mathbb{T}(Q)$ and another matrix $\mathbb{N}(Q)$, with elements 
$N_{s,s^\prime}(Q)=t(Q,s,s^\prime)e^{\beta\mu} \left [  \delta_{s,s^\prime+1}+ \delta_{s,s^\prime-1}\right ]$:
\begin{equation}\label{eq:density_classical}
\braket{m_j^2} = \frac{\int^{1/2}_{-1/2} dQ \,{\rm Tr}\,\left\{ \mathbb{N}(Q) \left [ \mathbb{T}(Q)\right ]^{L-1} \right\}}
{\int^{1/2}_{-1/2} dQ \,{\rm Tr}\,\left [ \mathbb{T}(Q)\right ]^L}  \;.
\end{equation}

Formally, both $\mathbb{T}(Q)$ and $\mathbb{N}(Q)$ are infinite matrices ($s$ and $s^\prime$ run from $-\infty$ to $+\infty$), but in the 
thermodynamic limit, two simplifications occur: first of all, evaluating the trace of $\left [ \mathbb{T}(Q)\right ]^L$ in the basis of the 
eigenvectors of $\mathbb{T}(Q)$, only the contribution of the eigenvector with the eigenvalue of largest absolute value remains (this is true 
if the spectrum of $\mathbb{T}(Q)$ is gapped, condition that is easily checked numerically); in addition, the integral over $Q$ can be evaluated 
by the steepest descent method and is dominated by the values $Q^{*}_{\sigma}$ maximizing the highest eigenvalue. Therefore, denoting the 
largest eigenvalue by $\kappa_0(Q)$ (satisfying the relation $\mathbb{T}(Q)\ket{u_0}=\kappa_0(Q)\ket{u_0}$), which is maximized by some 
values $Q^*_{\sigma}$, the following expressions hold in the thermodynamic limit:
\begin{equation}\label{eq:partitionthermodlim}
	\lim_{L\to\infty}\frac{{\log \cal Z}}{L} =\log \left [ \sum_{\sigma} \kappa_0(Q^*_{\sigma}) \right ] \;,
\end{equation}
\begin{equation}\label{eq:density_thermodlim}
\lim_{L\to\infty} \braket{m_j^2} = \frac{\sum_{\sigma} \braket{u_0 | \mathbb{N}(Q^*_{\sigma})|u_0}}{\sum_{\sigma} \kappa_0(Q^*_{\sigma})} \;.
\end{equation}
Density-density correlations $\braket{m_i m_j}$ or string correlations $\braket{m_i e^{i\pi \sum_{l=i+1}^{j-1} m_l} m_j}$ can be also evaluated 
by relations analogous to Eqs.~\eqref{eq:density_classical} and~\eqref{eq:density_thermodlim}. 

An intuitive picture of the origin of topological order in the one dimensional Coulomb gas can be gained by noting that the energy of two dipoles 
in the vacuum does not depend upon the distance between them, being lower for aligned dipoles (e.g., $\{0,0,+,-,0,0,+,-,0,0\}$ on $L=10$) rather 
than anti-aligned ones (e.g., $\{0,0,+,-,0,0,-,+,0,0\}$). As a results, there is a net dipole-dipole attraction, which favors the onset of the 
string order, identified by the string-order parameter: $\mathcal{S} = \lim_{|i-j|\to \infty} \braket{m_i e^{i\pi \sum_{l=i+1}^{j-1} m_l} m_j}$.

The zero-temperature properties of the classical Coulomb gas model are easily understood: for $\mu>1/4$, the lowest-energy configurations are 
$\{+,-,\dots,+,-\}$ and $\{-,+,\dots,-,+\}$, which correspond to perfect charge-density-wave (Ising antiferromagnetic) states in the charge (spin)
language. By contrast, for $\mu<1/4$ the state $\{0,0,\dots,0,0\}$ has the lowest energy, corresponding to the empty (nematic) state in the charge
(spin) language. At finite temperature, the charge-density-wave (antiferromagnetic) order does not survive, instead a state with finite string 
order emerges, signaling the presence of aligned dipoles in the system. The phase diagram is shown in Fig.~\ref{fig:phase_diag_classic_String},
displaying two different phases. For $\mu > \mu^{*}(T)$, there is a (dense) liquid of aligned dipoles, with {\em finite} string order; instead, 
for $\mu < \mu^{*}(T)$ a diluted gas of disordered dipoles, with no string order, stabilizes. In the same figure, we also report the values of 
the optimal parameters $\beta$ and $\mu$ of the variational wave function with the simplified Jastrow pseudo-potential of Eq.~\eqref{eq:vqsimple}, 
optimized for a chain of $L=200$ sites and values of $D/J$ ranging from $-0.5$ to $3.4$. From this picture, we clearly observe the mechanism by 
which the Jastrow wave function is able to represent both the nematic and the topological phases of the Heisenberg chain: while $D/J$ varies, the 
optimized parameters lead the variational wave function from one phase of the classical model to the other. 

The nature of the transition between the two phases may be investigated directly in the thermodynamic limit by inspecting the density and the
string parameter. In Fig.~\ref{fig:string_density} we show the average density $\braket{m_j^2}$ and the string order $\mathcal{S}$ as a function 
of $\mu$ for two different values of $\beta$. For low temperatures, there is a jump in both the density and string order, indicating that the 
transition between the dense and diluted phases is first order. In contrast, at higher temperatures, both quantities are continuous, the string-order 
parameter being zero on the diluted region and finite in the dense one. In this case, a second-order phase transition can be identified. The change 
from first- to second-order nature is estimated at $\beta \approx 2.5$, as shown in Fig.~\ref{fig:string_density}.

\subsection{The ground-state phase diagram of the Heisenberg model}\label{sec:fullphase}

In the previous subsections, we have shown how a simple Jastrow wave function is able to capture the correct physical content both in the 
topological (Haldane) and trivial (nematic) phases, through a long-range pseudo-potential $v_{r}$. A simple parametrization gives important 
insights on the origin of the two phases, with and without string order. Still, the optimal form of $v_{r}$ deviates from the parametrization of 
Eq.~\eqref{eq:vqsimple}. From a quantitative point of view, an explicit nearest-neighbor term is relevant to improve the variational energy, even 
well inside the gapped regimes. In Table~\ref{tab1}, we report the results obtained by using both the optimal Jastrow state and its simplified 
version, given by the parametrization of Eq.~\eqref{eq:vqsimple}.

\begin{table}
\caption{\label{tab1}
Total energies of the simplified (second column) and  optimized (third column) Jastrow functions for different values of $D/J$. The simplified 
Jastrow pseudo-potential is given by Eq.~\eqref{eq:vqsimple}. The relative error is also reported in the forth column, where 
$\Delta E=E_{\mathrm{opt}}-E_{\mathrm{simpl}}$. The size of the cluster is $L=200$.}
\begin{tabular}{cccc}
\hline
$D/J$ & $E_{\mathrm{simpl}}/J$ & $E_{\mathrm{opt}}/J$ & $\Delta E/E_{\mathrm{opt}}$ \\
\hline \hline
0.0  & -272.11(8) & -279.20(2) & 3.64(3) \% \\
0.3  & -235.14(7) & -241.36(7) & 2.58(4) \% \\
0.5  & -212.2(1)  & -218.51(9) & 2.89(6) \% \\
0.8  & -181.08(8) & -188.87(2) & 4.12(4) \% \\
1.0  & -163.96(7) & -170.89(1) & 4.06(4) \% \\
1.5  & -130.30(7) & -132.77(3) & 1.86(5) \% \\
2.0  & -105.04(4) & -105.99(3) & 0.90(5) \% \\
2.5  & -86.39(3)  & -86.76(3)  & 0.43(5) \% \\
3.0  & -72.63(2)  & -72.78(2)  & 0.21(5) \% \\
\hline \hline
\end{tabular}
\end{table}

\begin{figure}
\includegraphics[width=\columnwidth]{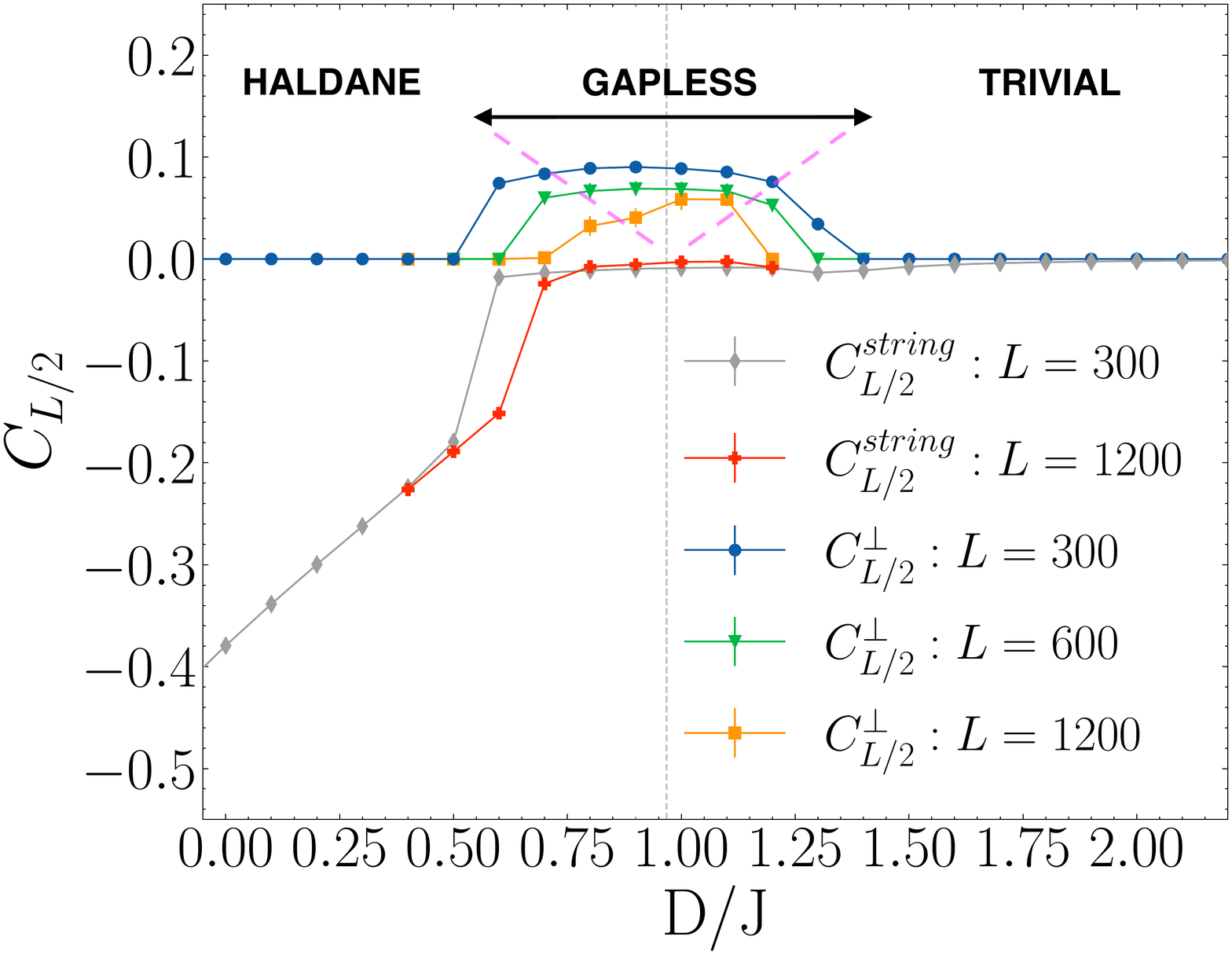}
\caption{\label{fig:orderparam} 
Order parameters for the Heisenberg chain with single-ion anisotropy $D/J$, for different sizes $L$.}
\end{figure}

\begin{figure}
\includegraphics[width=\columnwidth]{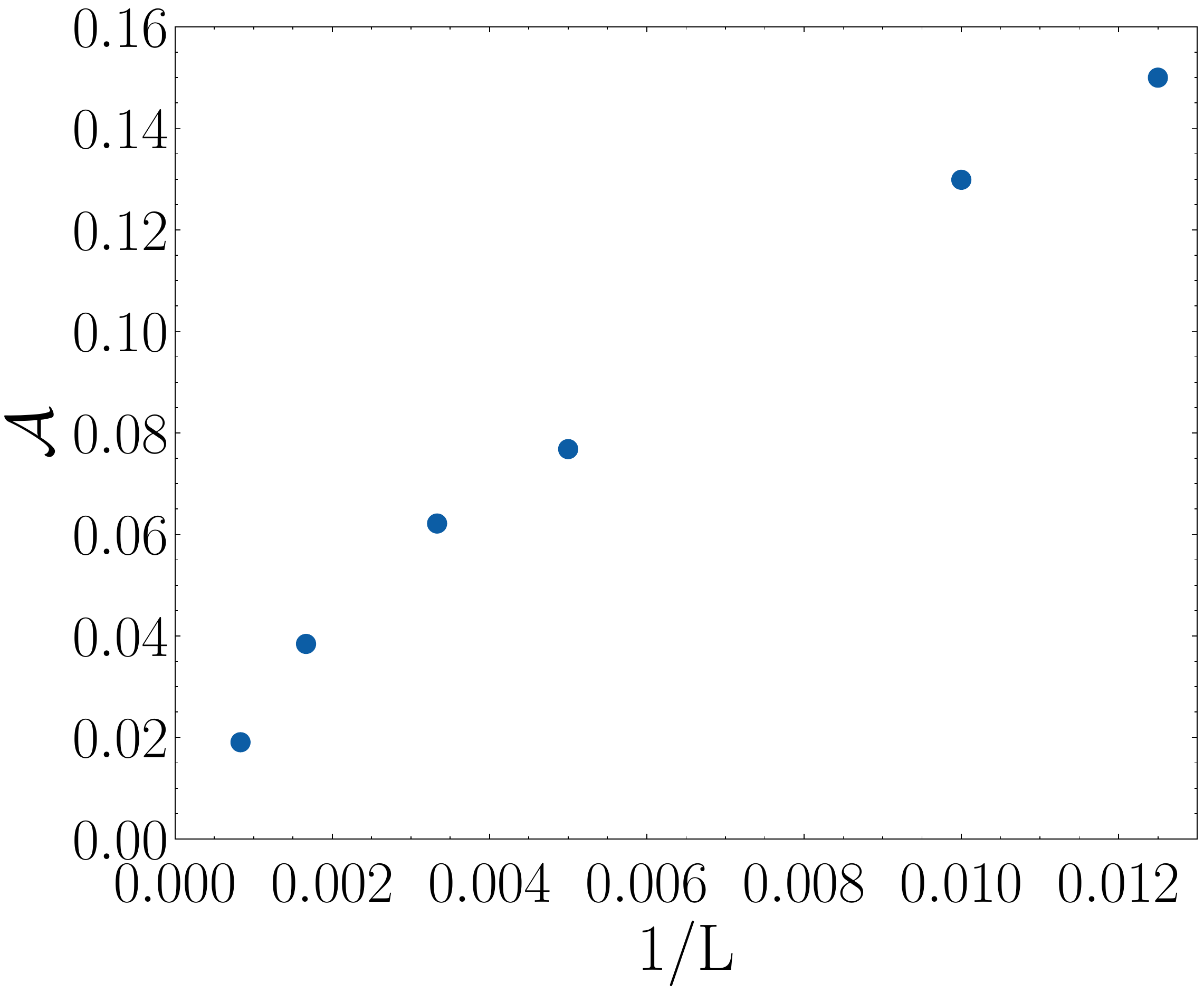}
\caption{\label{fig:area} 
Area $\mathcal{A}$ enclosed by the curve $C^{\perp}_{L/2}$ for different sizes $L$ plotted against $1/L$. The biggest size we consider is $L=1200$ 
and the area is computed from the curve $C^{\perp}_{L/2}$ sampled on a grid of $D$-points with spacing $0.1J$: the area goes to zero in the 
thermodynamic limit.}
\end{figure}

Most importantly, the simple parametrization does not reproduce the correct small-$q$ behavior of the fully-optimized pseudo-potential when the 
the transition between topological and trivial phases is approached and the cluster size $L$ is not sufficiently large. Indeed, the transition 
point is expected to be gapless with power-law correlations~\cite{hu2011}, which needs a less singular Jastrow factor with $v_{q} \approx 1/q$ 
for $q \to 0$~\cite{capello2005,capello2007}. On small clusters, this kind of pseudo-potential is obtained by a full optimization, thus broadening 
the critical point. Within this intermediate regime, the spin-spin correlations in the $x{-}y$ plane $C^{\perp}_{r}$ show a slow power-law decay, 
while there are no appreciable string correlations $C^{\mathrm{string}}_{r}$ at large distances. In Fig.~\ref{fig:orderparam}, we report the 
results for $C^{\perp}_{L/2}$ and $C^{\mathrm{string}}_{L/2}$, for different sizes $L$ and values of $D/J$. Well inside the stable phases the 
situation is clear: the Haldane phase has finite $C^{\mathrm{string}}_{L/2}$, while $C^{\perp}_{L/2}$ vanishes. Instead, for large values of 
$D/J$, the nematic phase settles down, where all correlations vanish. In the vicinity of the transition point, for small clusters the in-plane 
correlations $C^{\perp}_{L/2}$ are remarkably enhanced. Still, this is a finite-size effect, which disappears when $L$ becomes sufficiently large. 
Indeed, by increasing the cluster size $L$, the string-order parameter becomes progressively finite when approaching the transition, while 
$C^{\perp}_{L/2} \to 0$, see Fig.~\ref{fig:orderparam}. In order to better quantify this fact, we report in Fig.~\ref{fig:area} the area ${\cal A}$ 
below the curve $C^{\perp}_{L/2}$ in the intermediate phase against $1/L$: this area clearly shrinks to zero in the thermodynamic limit. We mention 
that a similar effect, with a broadening of the critical behavior is also seen in recent DMRG calculations~\cite{lambert2023}, suggesting that 
exceedingly large clusters are needed to detect the direct transition between topological and trivial phases.

\section{Conclusions}\label{sec:concl}

In conclusion, we numerically demonstrated that, despite its apparent simple form, the (two-body) Jastrow wave function is able to represent the 
Haldane phase and describe correctly the phase diagram of a Heisenberg chain with single-ion anisotropy. Therefore, this simple {\em Ansatz} can 
reproduce a rich and non-trivial physics, with the advantage of its compactness (it contains only $L/2$ parameters) and its direct physical 
interpretation. Furthermore, the Jastrow pseudo-potential assumes a remarkably simple form, which can be fitted by using a two-parameter function,
with an on-site term and a long-range tail. This approximation gives a straightforward interpretation thanks to a mapping between the quantum
expectation values over the variational state and a classical partition function. The phase diagram of the latter one shows two phases at low
temperatures (relevant for the optimized Jastrow pseudo-potential): one diluted gas of disordered dipoles (corresponding to the nematic phase 
of the Heisenberg model with large values of $D/J$) and one denser liquid of ordered dipoles (corresponding to the Haldane phase). Overall, this 
work further highlights the (often overlooked) merits of the Jastrow wave function, which features an optimal balance between accuracy and 
simplicity also in challenging many-body models.

\section*{Acknowledgements}
G.E.S. acknowledges financial support from PNRR MUR Project PE0000023-NQSTI.

\appendix
\begin{widetext}
\section{Details for the classical partition function}\label{app1}

Here, we provide the formal derivation of the classical partition function given in Eq.~\ref{eq:partitionfinal}.

Given a positive definite $L\times L$ matrix ${\bf A}$ and a $L$-dimensional vector ${\bf m}$, the basic identity (Hubbard-Stratonovich transformation) 
is
\begin{equation}
\int_{-\infty}^\infty \prod_j d\phi_j \, \exp \left \{ -\sum_{j,k} \phi_j A_{j,k} \phi_k + i \sum_{j} m_j \phi_j \right \}  =
\sqrt{\frac{\pi^{L}}{\det {\bf A}}} \exp \left \{-\frac{1}{4} \sum_{j,k} m_j (A^{-1})_{j,k} m_k \right \} \;.
\end{equation}
Then, by taking
\begin{equation}
(A^{-1})_{j,k} = \frac{2 \beta}{L} \left [ \left (d_{j,k}-\frac{L}{2}\right)^2 +\Lambda L \right ] \;,
\end{equation}
the partition function is written as:
\begin{equation}\label{eq:partition1}
{\cal Z} = \sqrt{\frac{\det {\bf A}}{\pi^{L}}} \int_{-\infty}^{\infty} \prod_j d\phi_j \, 
\exp \left \{-\sum_{j,k} \phi_j A_{j,k} \phi_k \right \} \prod_{j} \sum_{m=-1,0,+1} e^{\beta\mu m^2 + i m \phi_j} \;,
\end{equation}
where the matrix ${\bf A}$ can be easily found because it is diagonalized by Fourier transform:
\begin{equation}
A_{j,k} = \frac{1}{2\beta} \left [ \frac{1}{L} \left ( \Lambda L + \frac{L^2+2}{12}\right )^{-1} +
          \delta_{j,k} - \frac{1}{2} (\delta_{j,k+1} + \delta_{j,k-1})\right] \;,
\end{equation}
whose determinant is 
\begin{equation}
\det {\bf A}=\frac{2L^2}{U_0 (4\beta)^{L}} \;
\end{equation}
where $U_0=\Lambda L + \frac{L^2+2}{12}$, see Eq.~\eqref{eq:defvq}. Then, we define
\begin{equation}
\Gamma=\frac{1}{2\beta L U_0} \;,
\end{equation} 
such that $\det {\bf A}/\Gamma=L^3/(4\beta)^{L-1}$. 

The summation over $m$ in Eq.~\ref{eq:partition1} can be easily done. Then, by inserting the identity 
$\int_{-\infty}^{\infty} dx \delta(\sum_j\phi_j-x)=1$ and using the integral representation of the $\delta$-function 
$\delta(t) = \int_{-\infty}^{\infty} \frac{d\omega}{2\pi} e^{i\omega t}$, we get
\begin{equation}\label{eq:partition2}
{\cal Z} = \sqrt{\frac{L^3}{(4\pi\beta)^{L-1}}} \, \int_{-\infty}^{\infty} \frac{d\omega}{2\pi} \,e^{-\frac{\omega^2}{4\Gamma}}\,F(\omega)
\end{equation}
where
\begin{equation}
F(\omega) = \int_{-\infty}^{\infty} \prod_j d\phi_j \, \exp \left \{ -\frac{1}{4\beta} \sum_{j} (\phi_{j}-\phi_{j-1})^2 \right \}
\prod_{j} e^{i\omega\phi_j} \, \left [ 1 + 2e^{\beta\mu}\cos\phi_j \right ] \;.
\end{equation}
Now, the parameter $\Lambda$ appears only through $\Gamma$ ruling the amplitude of the Gaussian term in Eq.~\eqref{eq:partition2}. Charge neutrality 
is enforced when $\Lambda \to \infty$, i.e., for $\Gamma\to 0$. The function $F(\omega)$ must have a $\delta(\omega)$ contribution, in order for 
the partition function to attain a finite limit: if $F(\omega) \approx F_0 \,\delta(\omega)$ for small $\omega$, the partition function 
becomes proportional to the amplitude $F_0$. Therefore, our main task is the evaluation of the small $\omega$ behavior of $F(\omega)$. 

The partition function can be written in a more symmetric way, by defining the operator~\cite{demery2012}:
\begin{equation}
K_\omega(x,x^\prime) = \frac{1}{\sqrt{\pi\beta}}\,\int_{-\infty}^{\infty} dy \, e^{-\frac{(x-y)^2}{2\beta}}
e^{i\omega y} \,\left [ 1 + 2e^{\beta\mu} \cos y \right ] e^{-\frac{(y-x^\prime)^2}{2\beta}} \;.
\end{equation}
It is easy to check that $F(\omega)$ is just the trace of the product of $L$ of these integral operators:
\begin{equation}\label{eq:fomega}
F(\omega)=\int d\phi_1\dots d\phi_L  \,K_\omega(\phi_1,\phi_2) \,K_\omega(\phi_2,\phi_3)\dots K_\omega(\phi_L,\phi_1) \;,
\end{equation}
which can be also written as
\begin{equation}
F(\omega) = \int d\phi_1\,d\phi_1^\prime\dots d\phi_L\,d\phi_L^\prime  \,K_\omega(\phi_1,\phi_1^\prime)
\,\delta(\phi_1^\prime-\phi_2)\,K_\omega(\phi_2,\phi_2^\prime)\,\delta(\phi_2^\prime-\phi_3)\dots K_\omega(\phi_L,\phi_L^\prime)
\delta(\phi_L^\prime-\phi_1) \;.
\end{equation}
By using the integral representation of the $\delta$-function
\begin{equation}
\delta(\phi^\prime-\phi) = \int_{-\infty}^\infty \frac{dq}{2\pi} \, e^{iq(\phi^\prime-\phi)} \;,
\end{equation}
the integrals over $\phi$ and $\phi^\prime$ can now be performed with the result:
\begin{eqnarray}
K_\omega(p,q) &=& \int d\phi \,d\phi^\prime \,e^{-ip\phi+iq\phi^\prime} \,K_\omega(\phi,\phi^\prime)= \nonumber \\
              &=& 4\pi\,\sqrt{\pi\beta}\, e^{-\frac{\beta}{2} (q^2+p^2)} \,
\left [\delta(q-p+\omega)+ e^{\beta\mu} \delta(q-p+\omega+1)+ e^{\beta\mu} \delta(q-p+\omega-1)\right ] \;.
\label{eq:kpq}
\end{eqnarray}
In terms of this quantity:
\begin{equation}
F(\omega) = \int\frac{dq_1}{2\pi} \dots \frac{dq_L}{2\pi}\, K_\omega(q_1,q_2)\dots K_\omega(q_L,q_1) \;.
\end{equation}
Due to the specific form of $K_\omega(p,q)$ [see Eq.~\eqref{eq:kpq}], for each choice of $q_1$ there is a unique choice of $q_2$ up to $q_L$
and only the integral over $q_1$ survives. Fixing an arbitrary value of $q_1$, the $\delta$-functions present in $K_\omega(q_1,q_2)$ force 
$q_2=q_1-\omega+n_2$ with $n_2=0,\pm 1$. Then $q_3=q_1-2\omega + n_3$ with $n_3-n_2=0,\pm 1$ and so on. Then all $q_j$ have the form 
$q_1-(j-1)\omega+n_j$ and the last term $K_\omega(q_L,q_1)$ then gives $q_1 = q_L -\omega + (n_{L+1}-n_L)$, i.e., $q_1=q_1-L\omega+n_{L+1}$, 
meaning that $n_{L+1}=0$ and the function $F(\omega)$ is proportional to a $\delta(L\omega)$. Setting $q_1=Q+n_1$ with $n_1$ an arbitrary integer 
and $|Q|< \frac{1}{2}$, the integral over $q_1$ splits into an integral over $Q$ and a sum over $n_1$, while the condition $n_{L+1}=0$ implies 
that $F(\omega)$ is equal to 
\begin{equation} 
F(\omega) = \sqrt{(4 \pi\beta)^{L}} \, \delta(\omega L) \,\int^{\frac{1}{2}}_{-\frac{1}{2}} dQ \,{\rm Tr}\,\left [ \mathbb{T}(Q)\right ]^L \;,
\end{equation}
which, inserted in Eq.~\eqref{eq:partition2} gives the final expression in Eq.~\eqref{eq:partitionfinal}.
\end{widetext}

\bibliography{references}

\end{document}